\documentstyle[12pt]{article}
\global\arraycolsep=1pt
\begin{document}

\newcommand{\ba}{\begin{eqnarray}}
\newcommand{\ea}{\end{eqnarray}}
\newcommand{\be}{\begin{equation}}
\newcommand{\ee}{\end{equation}}
\newcommand{\barint}{-\hskip -11pt\int}

\begin{titlepage}
\begin{flushright}
YITP-97/044 \\
LPTENS-97/44 \\
October 1997
\end{flushright}

\begin{center}
{\bf A MATRIX MODEL SOLUTION OF HIROTA EQUATION\footnote{
written for the proceedings of the NATO Advanced Research Workshop
``New Developments in Field Theory'', Zakopane, Poland, June 14-20, 1997}}
\vskip 0.6 in
{\bf Vladimir A. Kazakov}
\vskip 0.6 in
Laboratoire de Physique Th\'eorique de
l'\'Ecole Normale Sup\'erieure \footnote{
Unit\'e Propre du
Centre National de la Recherche Scientifique,
associ\'ee \`a l'\'Ecole Normale Sup\'erieure et \`a
l'Universit\'e de Paris-Sud}\\
24 rue Lhomond, F-75231 Paris Cedex 05, France\\
and\\
 Yukawa Institute for Theoretical Physics\\
Kyoto University, Sakyo-ku, Kyoto 606-01, Japan

\end{center}

\vskip 0.6 in
\begin{center}
{\bf ABSTRACT}
\end{center}
We present a hermitian matrix chain representation of the general
solution of the Hirota bilinear difference equation of three
variables. In the large N limit this matrix model provides some
explicit particular solutions of continuous differential Hirota
equation of three variables. A relation of this representation to the
eigenvalues of transfer matrices of 2D quantum integrable models is
discussed.

\end{titlepage}

\section{INTRODUCTION}

The Hirota bilinear equations (HE) [\cite{HIR}] provide, may be, the
most general view on the world of exactly solvable models, from
integrable hierarchies of differential and difference equations
equations, like KdV equations or Toda chains, to, as it was recently
shown, the transfer matrices of 2D models of statistical mechanics and
quantum field theory [\cite{KLPI,KUN,KLWZ}] integrable by the Bethe
ansatz (BA) techniques.  Indeed, in the last case, it was shown that
the eigenvalues of transfer matrices must obey the general HE with 3
discrete variables corresponding to the rapidity, rank and level of a
representation with rectangular Young tableaux, used in the fusion
procedure. The Thermodynamical Bethe Ansatz (TBA) equations discovered
by Lee and Yang and widely used in the last years for the description
of thermal properties and finite size effects of 2D integrable models
follow almost directly from this HE [\cite{ALZ}].

HE has its own history in connection with the matrix models (MM).  The
general solitonic solutions of HE [\cite{MI}] bear a striking
resemblance with the matrix models with logarithmic potentials. The
discovery of double scaling limit in the matrix models
[\cite{BK,GM,DS}] (corresponding to the  big size of  matrices and a special
tuning of potentials) showed that the matrix models are closely
related to KdV and KP hierarchies of integrable differential equations
[\cite{DOU}]. The study following it provided even more general examples
of this correspondence: the multi matrix models before taking any
large N limit appeared to be related the classical Toda chains
[\cite{MMM}].  Another manifestation of these connections are the
Schwinger-Dyson equation for MM which can be written in terms of
Virasoro constraints [\cite{YUT,KMMM}] (see [\cite{KOS}] for a modern
account of this approach).

We would also recall one rather mysterious coincidence: in [\cite{KAKO}]
an open string amplitude for the (1+1)D string with mixed
(Dirichlet-Neumann) boundary conditions appeared to have the same form
as the Sine-Gordon S-matrix.

All this suggests that matrix models could have something to do with
the quantum 2D integrable models and the way from one to another might
go through HE.

In this paper we will propose a matrix chain representation of the
 general solution of Hirota difference equation. We will use for that
 the analogy between the so called Bazhanov-Reshetikhin determinant
 representation [\cite{BR}] of the transfer-matrix eigenvalues of
 integrable models (which obeys the HE) and the determinant
 representation of matrix chain, in terms of eigenvalues of the
 matrices. The potentials acting at every site of the chain will
 depend on the eigenvalues of the matrices and on the coordinate of
 the site. Hence the arbitrary potential is a function of 2 variables
 which is in general enough to parameterize any solution of a
 difference equation of 3 variables.

The solutions of HE relevant to TBA must obey very special boundary
conditions. It can include, for example, the condition on the maximal
possible size of the fusion representation (reflecting the invariance
of the initial model with respect to some continuous symmetry), and
Lorentz invariance of the spectrum of physical particles emerging
on the top of physical (dressed) vacuum.

It is not easy to extract the physical information from HE or the
corresponding TBA equations as it is not easy to solve the nonlinear
difference or integral equations with specific boundary
conditions. The results obtained on this way are quite limited: they
mostly concern the calculations of central charges and dimensions of
operators in the conformal (ultraviolet) limits and various
asymptotic expansions corresponding to high energies (see for example
[\cite{BR, ZHPI, BLZ}]). It even appears to be difficult to reproduce
the first loop calculation for the asymptotically free models in
finite temperature, apart from some simplified models [\cite{ALZ2}]),
although numerically the TBA equations work quite well. Another
challenge is to find the planar (large N) limit of such an interesting
2D quantum field theory as the principal chiral field. The model was
formally integrated by the BA approach in [\cite{WPO,PW}]. It has been
solved rather explicitly in the large N limit in case of zero
temperature and arbitrarily big external field [\cite{FKW1,FKW2}], but
the attempts to generalize it to finite temperatures were not
successful.

On the other hand, the general solution of HE follows from its
integrability and can be represented in terms of a $\tau$-function. In
case of a general difference HE it coincides with the determinant
representation of Bazhanov-Reshetikhin. It might be a good idea to
use this representation, and hence our matrix chain representation, to
get some hand on HE and TBA. Of course, for a finite rank N of matrices
(and hence finite dimensional groups of symmetry of corresponding
models) our representation hardly could offer some breakthrough. But
for big N we can try to use the machinery of the matrix models (like
orthogonal polynomials, character expansions and various saddle point
techniques) to calculate the corresponding infinite determinant.

We were not able to find any solutions of HE satisfying correct
TBA-like boundary conditions. A natural way to impose these boundary
conditions is the most important drawback of our
representation. Leaving it to future studies we propose here some
particular solutions of continuous (differential) Hirota equations
which describe our matrix chain in the large N limit. They correspond
to a particular choice of parameters (potentials) of the chain.

In the next section we will briefly review how the Hirota equation is
connected with the integrable models of the 2D quantum field
theory. The continuous differential version of them will be presented.

In   section 3 we will propose the matrix chain representation of
the difference Hirota equation based on the Bazhanov-Reshetikhin
determinant representation of the fusion rules for transfer-matrices.

In section 4 we will present some particular examples of solution of
the continuous differential HE, given by the one matrix model and the
matrix oscillator with the specific boundary conditions.

In section 5 we will sketch out a general solution of the differential
HE for an arbitrary time dependent matrix chain potential and consider
a more explicit solution for the particular case of time independent
potential.

The last section will be devoted to conclusions and prospects.

\section{TBA, FUSION RULES \\
AND HIROTA DIFFERENCE EQUATION}

To set a more physical background for our construction let us briefly
review how the HE appears from the Bethe ansatz. We will mostly follow
in this section the framework and the notations of
[\cite{KLWZ},\cite{ZAB}].

The transfer-matrix of an integrable 2D model with periodic boundary
conditions depends (apart from a number of fixed parameters, like
volume, temperature or the anisotropy $q$) on three variables:
rapidity $u=i\theta$, rank (``color'') $a$ and level (``string''
length) $s$.  The variables a and s have the meaning of a
representation of elementary spins filling the bare vacuum of the
model, given by the rectangular Young tableau of the size $a\times s$.
The corresponding transfer matrix is called $\hat T^a_s(u)$.

The integrability imposes the commutativity of transfer-matrices for
different values of all three variables playing thus the role of
spectral parameters:

\be
[\hat T^a_s(u),\hat T^{a'}_{s'}(u')]=0
\label{COM}
\ee

It follows from (\ref{COM}) that we can always work with the eigenvalues
$T^a_s(u)$ instead of the transfer-matrix itself and view them
as usual functions.

The transfer-matrices, as well as their eigenvalues, obey a set of
relations known as fusion rules, originally found as the relations
between S-matrices for particles with different spins in integrable
QFT. They can be summarized in the so called Bazhanov-Reshetikhin
formula [\cite{BR}] (BR) presenting the function of three variables
$T^a_s(u)$ in terms of the function of only two variables $T^1_s(u)$:
\be
T^a_s(u)=det_{1\le i,j \le a} T^1_{s+i-j}(u+i+j+a)
\label{BR}
\ee

Actually, there exists a more general BR formula, expressing the
transfer-matrix eigenvalue of any skew representation $h/h'$ through
$T^1_s(u)$:
\be
T_{h/h'}(u)=det_{1\le i,j \le a} T^1_{h_i-h'_j}(u+h_i+h'_j )
\label{BRG}
\ee
where $h_i=m_i+a-i$ and $h'_i=m'_i+a-i$ are the so called shifted
highest weight components of two representations $R$ and $R'$ of GL(N)
characterized by the usual highest weight components $R=(m_1,...m_a)$
and $R'=(m'_1,...m'_a)$, so that they obey the inequalities
$h_i<h_{i-1}$, $h'_i<h'_i$ and also $h'_i<h_i$. But the
transfer-matrices with rectangular Young tableaux play an exceptional
role since they obey a closed set of fusion rules given by the
difference Hirota equation:
\be
 T^a_s(u+1) T^a_s(u-1)- T^a_{s+1}(u) T^a_{s-1}(u) =  T^{a+1}_s(u)
T^{a-1}_s(u)
\label{HE}
\ee
It follows directly from  (\ref{BR}) in virtue of the Jacobi identity
for  determinants. It contains  little information since it is true
for any function of two variables $T^1_s(u)$ in (\ref{BR}). To specify
it further to some particular integrable model we have to impose some
boundary conditions on solutions of the eq. (\ref{HE}).

One of these conditions specifies the group of symmetry of the
model. To make it, say, SU(N) (or $A_{N-1}$ in terms of underlying
algebra) we put:

\be
 T^a_s(u)=0, \ \ \ \ \ \ for \ \ \ \ a < 0 \ \ and  \ \ \ \ a > N
\label{BKN}
\ee

It is not enough since it leaves us with an infinite discrete set of
possible solutions (like in quantum mechanics, fixing the boundary
conditions on a wave function we are still left with infinitely many
wave functions corresponding to different energy levels).  We have to
specify some analytical properties of solutions.

There are two ways to do it in the case of BA.

One is related to the so called bare BA where one specifies $T^0_s(u)$
and $T^N_s(u)$ to be some given polynomials in the variable $u$ whose
zeroes specify completely a model, where as the functions $T^a_s(u),
\ \ for \ \ 1 \le a \le N-1$ are polynomials whose zeros we have to find.
The details of the analyticity conditions for the bare BA can be found
for example in (\cite{KLWZ}).

Another way to fix analytical properties corresponds to the dressed BA
where the elementary excitations are already the real physical
particles. To precise them let us derive from (\ref{HE}) the TBA
equations. For that we introduce the function:
\be
Y^a_s(u)= { T^a_{s+1}(u) T^a_{s-1}(u) \over  T^{a+1}_s(u)
T^{a-1}_s(u)}
\label{Y}
\ee
which, in virtue of (\ref{HE}), satisfies the equation sometimes
called Y-system [\cite{ALZ}]:

\be
 {Y^a_s(u+1) Y^a_s(u-1) \over Y^{a+1}_s(u) Y^{a-1}_s(u)}
= {\big[1+Y^a_{s+1}(u)\big] \big[1+Y^a_{s-1}(u)\big]
 \over
\big[1+Y^{a+1}_s(u)\big] \big[1+Y^{a-1}_s(u)\big]}
\label{YS}
\ee

Note that this system is symmetric under the change: $Y \rightarrow
Y^{-1}, \ \ a \rightarrow s, \ \ s \rightarrow a$ (rank-level duality).

To make it a little bit more symmetric let us introduce the functions:
\be
U^a_s(u)= 1+Y^a_s(u), \ \ \ \ \tilde U^a_s(u)= 1+ {1 \over Y^a_s(u)}
\label{UU}
\ee

Then the eq. (\ref{YS}) can be rewritten as
\be
{ U^a_s(u+1) U^a_s(u-1) \over  U^a_{s+1}(u)
U^a_{s+1}(u)}={\tilde U^a_s(u+1) \tilde U^a_s(u-1) \over  \tilde U^{a+1}_s(u)
\tilde U^{a-1}_s(u)}
\label{UUS}
\ee

Taking logarithm of both sides of this equation and applying the
operator
\be
\int_{-\infty}^{\infty} d\theta'{1 \over
\cosh{\pi \over 2}(\theta-\theta')} \cdots
\label{OPR}
\ee
where $\theta=i u$  we obtain:
\be
C_{ss'}(\theta)*\ln \big(1+Y^a_{s'}(\theta)\big)=
C^{aa'}(\theta)*\ln \big(1+{1 \over Y^a_{s'}(\theta)}\big)
\label{CC}
\ee
where we introduced the so called ``baxterized'' Cartan matrices:
\be
C^{aa'}(\theta)=\delta_{a,a'}- {1 \over 2\cosh{\pi \over 2}(\theta)}
\big(\delta_{a,a'+1}+\delta_{a+1,a'}\big)
\label{CM}
\ee
and similarly for $C_{ss'}$. The $*$ sign defines the usual
convolution operation: $f(\theta)*g(\theta)=\int_{-\infty}^{\infty}
d\theta' f(\theta-\theta') g(\theta')$.

Let us now act on both sides of eq. (\ref{CC}) by the operator inverse
to (\ref{CM}):
\be
A_N^{aa'}= C^{-1}_{aa'}(\theta) =
{2 \over \pi} \int_0^{\infty} dp \cos(p\theta) \coth(p)
{\sinh[(N-max(aa'))p] \sinh[min(aa')p] \over \sinh(p N) }
\label{CINV}
\ee
Note that this inverse is respecting the boundary conditions
restricting the values of $a,a'$ to  $1 \le a,a' \le N-1$.

Note also that the operator $C^{aa'}$ has zero modes:
\be
C^{aa'}(\theta)*\sigma(s)m_0\sin(a'\pi k/N)\cosh(\theta \pi k/N)=0
\label{ZM}
\ee
for any integer k and any function $\sigma(s)$. So in acting by
(\ref{CINV}) on both sides of (\ref{CC}) we might be obliged to add
one of zero modes. The choice of zero mode and the function
$\sigma(s)$ defines completely the boundary conditions and hence the
model. If we want to respect the 1+1 dimensional relativistic
invariance we can add the zero mode with $k=1$, since only it will
lead to the relativistic spectrum of energies elementary excitations
(which are described by this zero mode) of a type:
\be
 \sigma (s) m_0\sin(a\pi/N)\cosh(\theta \pi/N)
\label{ES}
\ee
It gives a typical mass and energy spectrum of physical particles for
integrable relativistic models of 2D QFT. The choice of $\sigma(s)$
and the range of $s$ define a particular relativistic model. For
example, for $\sigma(s)=\delta_{s,1}, \ s \ge 1$ corresponds to the
chiral Gross-Neveu model, whereas $\sigma(s)=\delta_{s,0}, \ -\infty
\le s \le \infty$ corresponds to the principal chiral field (PCF) with
the SU(N) symmetry.

With all these settings the final TBA equation (or  similar
equations for the ground state of the finite length system with the
periodic boundary conditions) takes a familiar form (say, for the PCF):
\be
A^{aa'}(\theta)*C_{ss'}(\theta)*\ln \big(1+Y^a_{s'}(u)\big)-
\ln \big(1+{1 \over Y^a_{s'}(u)}\big)=
\delta_{s,0}m_0\sin(a\pi/N)\cosh(\theta \pi/N)
\label{TBA}
\ee
where $\epsilon(x,y,\tau)=\log Y^a_s(u)$ plays the role of the energy
density of the excitations characterized by the rank $a$ and level
$s$.

At the end of this section let us comment on the large N limit of the
TBA equations. It is not obvious to us how to simplify the eq. (\ref{TBA})
in this limit, but the difference equation equation (\ref{YS}) after
introducing the rescaled continuous variables
\be
\tau=u/N, \ \ \ \eta =s/N, \ \ \ \nu=a/N
\label{NVR}
\ee
becomes a second order differential equation for the quantity
$\epsilon(\eta,\nu,\tau)=\log Y^a_s(u)$ \footnote{I
thank P. Zinn-Justin for this comment}:
\be
(\partial^2_\nu - \partial^2_\tau ) \epsilon=
(\partial^2_\nu - \partial^2_\eta ) \ln\big(1+\exp\epsilon\big)
\label{DE}
\ee
This is an integrable classical equation, as it is a consequence of
the general HE and the determinant representation
(\ref{BR}) for its solution, although the determinant becomes
functional in the large N limit.

Let us give also another form of this equation in terms of new
 variables $l,m$ and $a$ ( and the corresponding rescaled variables)
$\lambda$, $\mu$ and $\nu$  defined through the
original variables $u,s$ and $a$ as follows:
\be
\lambda = {l \over N}={u+s+a \over 2N},
\ \ \ \mu = {m \over N}={u-s+a \over 2N},     \ \ \
\nu = {a \over N}
\label{PAR}
\ee
Note that $\lambda$ and $\mu$ play the role of ``light-cone'' variables.

In their terms we can represent the function $\epsilon=\ln Y$
in the large N limit as
\be
\epsilon=-\ln {1 \over \exp F''_{\lambda \mu} -1}
\label{YF}
\ee
where
\be
F=lim_{N \rightarrow \infty} {1\over N^2} \log T.
\label{SMF}
\ee

The continuous HE (\ref{DE}) for this function looks in the new
variables as
\be
\partial_\lambda \partial_\mu \ln (\exp [-F''_{\lambda \mu}] -1)=
\partial_\nu (\partial_\nu- \partial_\lambda- \partial_\mu)
F''_{\lambda \mu}]
\label{NHE}
\ee
This form of HE will be useful for our matrix model representation of
its solution.

\section{MATRIX MODEL CHAIN AS SOLUTION OF THE BILINEAR DIFFERENCE \\
HIROTA EQUATION}

In this section we propose to parametrize the general solution of  HE
(\ref{HE}) by means of  so called matrix chain integral - a matrix
model widely used and investigated in the literature (see [\cite{KD1}] for
the details).

Let us define the following Green's function:
\be
K_a(T,\phi(0),\phi(T)) = \int \prod_{k=1}^{T-1} d^{a^2} \phi(k)
\exp  tr\big[\sum_{k=0}^{T-1} \phi(k) \phi(k+1) - \sum_{k=0}^T
V(k,\phi_k)\big]
\label{MC}
\ee
where $\phi(k)$ - are $a\times a$ hermitian matrices, each
corresponding to its site $k$ of the chain, with matrix elements
$\phi_{ij}(k)$, $i,j=1,2,\cdots,N$, $k=0,1,2,\cdots,T$. Two matrices
at the ends of the chain are fixed.

The continuous analogue of the chain would be just the matrix quantum
mechanical Green's function on the interval of time $(0,T)$:
\be
K_a(T,\phi(0),\Phi(T)) = \int D^{a^2} \phi(t)
\exp-tr\int_0^T dt [{1 \over 2} \dot \phi^2 +V(t,\phi)]
\label{MQM}
\ee
with the end point values of the matrix fields also fixed. The matrix
quantum mechanics was introduced and solved in (\cite{BIPZ}).

Let us now define  a new quantity:
\be
Z_a(l,m)= \int d^{a^2} \phi(0) \int d^{a^2} \phi(T) (\det\phi(0))^l
K_a(T,\phi(0),\phi(T)) (\det\phi(T))^m
\label{TAU}
\ee
where $K$ could be any of both Green's functions (\ref{MC}) or
(\ref{MQM}). We have added two logarithmic potentials at the ends of
the chain to describe the dependence on $l$ and $m$ introduced by
(\ref{PAR}).

We claim that the function of three discrete variables
\be
T^a_s(u)=Z_a({u+s+a \over 2},{u-s+a \over 2})
\label{SOL}
\ee
obeys the HE (\ref{HE}). More than that,  it gives the most general
solution of HE parameterized by the function of 2 variables - the
potential V(t,x).

The proof goes as follows. If we start, say, from the discrete version
we can diagonalize each of the matrices in the chain by the unitary
rotation:
\be
(\phi_a)_{ij}=\sum_k (\Omega^+_a)_{ik}(z_a)_k(\Omega_a)_{kj}
\label{UR}
\ee
and integrate over the relative ``angles'' between two consecutive
matrices in the chain by means of the Itzykson-Zuber-Harish-Chandra
formula. This concerns only the first term in the exponent in the
r.h.s of (\ref{MC}). The potentials, including two determinants at the
ends, depend only on the eigenvalues. The overall result after the
angular integration will be (see (\cite{KD1}) for the details of this
calculation):
\be
Z_a(l,m) = \det_{1 \le i,j \le a} \int dp \int dq
K(T,p,q)p^{l+i-1} q^{m+j-1}
\label{DR}
\ee
where
\be
K(T,p,q) = \int \prod_{k=1}^{T-1} d z(k)
\exp \{\sum_{k=0}^{T-1} z(k) z(k+1) - \sum_{k=0}^T  V(k,z(k))\}
\label{VC}
\ee
for the discrete chain, or
\be
K(T,p,q) = \int D z(t)
\exp-\{\int_0^T dt[{\dot z^2\over 2}+V(t,z)]\}
\label{QM}
\ee
for the continuous quantum mechanics, where $z(0)=p, z(T)=q$.

Now we see that due to the determinant representation (\ref{DR}) the
function $T^a_s(u)$ defined by the eq. (\ref{SOL}) has the same
determinant form as the BR formula  (\ref{BR}) and
hence it obeys the HE (\ref{HE}), if we identify
\be
T^1_s(u) =  \int dp \int dq
K(T,p,q)p^{{u+s \over 2}} q^{{u-s \over 2}}
\label{T1}
\ee
The formula (\ref{DR})  generally defines an arbitrary function of two
variables $s$ and $u$. It is clear from the fact that it is just the
Mellin transform of an arbitrary function (\ref{QM}) (or (\ref{VC})) in two
variables, which is in our case
the Greens function of a quantum mechanical particle in an arbitrary
time and space dependent potential. This potential obviously gives
enough of freedom to define $K(T,p,q)$ as an arbitrary function of two
variables $p$ and $q$. So we proved (or at least made rather obvious)
 our statement about the generality of this representation of solution
of the HE.

We can provide a more general matrix integral giving the
parameterization of the most general BR formula (\ref{BRG}):
\be
T_{h/h'}= \int d^{a^2} \phi_0 \int d^{a^2} \phi_T \chi_{[h]}(\phi_T)
K_a(T,\phi_0,\phi_T) \chi_{[h']}(\phi_T)
\label{TAUG}
\ee
where $\chi_{[h]}(\phi)$ is the GL(a)  character of the representation
characterized by the highest weight $[h]$. It can be easily proved
when written in terms of eigenvalues with the use of
the Weyl formula for characters.

For any finite N all this seems to be on the edge of triviality: we
just defined in a sophisticated way an arbitrary function of two
variables and built from it the necessary determinant. Naturally, we
don't expect this representation to be of big use for a finite N.  The
boundary conditions of TBA will be as difficult to satisfy as before.
What is our major hope is the large N limit of this matrix model which
should correspond to the large N limit of the integrable models of the
type of principal chiral field. In this case the determinant in the BR
formula is essentially functional, and the matrix models give a rich
variety of methods for the calculation of such determinants. That why
our strategy will be the following: we investigate the matrix
integrals of the type (\ref{TAU}) in the large N limit for various
potentials and look for physically interesting regimes. Things might
become much more universal in the large N limit, and it could exist a
classification of interesting regimes, like it was done for the
multi-critical points in the matrix models. This paper represents of
course only a few modest steps in this direction.

Let us make an important remark concerning the large N limit of the
representation (\ref{DR}): the Y variable introduced in the previous
section and presented by the formula (\ref{YF}) in the large N limit
(with the matrix chain partition function $Z$ instead of the
transfer-matrix $T$) obeys the differential equation (\ref{DE}) or, in
new variables (\ref{PAR}), (\ref{NHE}). It is almost clear from our
definitions and it will be demonstrated in the following sections.


Another  more formal but interesting application of this method could be the
search for  new solutions in the integrable equations of the type
(\ref{DE}). In the next section we will show some particular examples
of it.

\section{EXAMPLES OF SOLUTIONS \\
OF THE CONTINUOUS HIROTA EQUATION}

We will demonstrate here on some examples limited to particular
choices of the matrix chain potentials how this relation between the
HE and MM works.

\subsection{One matrix model and GL(n) character}

Let us start from the one matrix model partition function
with an extra logarithmic potential
\be
Z_a(l)= \int d^{a^2}\phi (\det \phi)^l \exp-N tr V(\phi)
\label{1MM}
\ee
which, after going to the eigenvalue representation, becomes
\be
Z_a(l) = \det_{1 \le i,j \le a} \int dp \ p^{l+i+j-2}\exp-N V(x)
\label{D1MM}
\ee
It is well know that if one chooses:
\be
\exp-V(x)=\prod_{k=1}^a(b_k-x)^{-1}
\label{MV}
\ee
and performs the integral in (\ref{D1MM}) along the contour encircling
all these poles one will identify this partition function with the
GL(a) character $\chi^a_l(b)$ of the $a \times l$ rectangular Young tableau
given by the Weyl determinant formula. So, much of our next formulas is
valid for the characters as well.

It is easy to see from the Jacobi identity for determinants that the
function
\be
t^a(l)=Z_a(a-l)
\label{HF}
\ee
 satisfies a simplified version of the general HE
(\ref{HE}):
\be
 t^a(l+1) t^a(l-1)- t^{a+1}(l) t^{a-1}(l)=[t^{a}(l)]^2
\label{CHE}
\ee
Introducing the variable
\be
Y^a(l)= { t^{a+1}(l) t^{a-1}(l) \over  [t^a(l)]^2 }
\label{SY}
\ee
we obtain from (\ref{CHE}) a simplified version of the general
Y-system (\ref{YS}):
\be
 {Y^a(l+1) Y^a(l-1) \over [Y_a(l)]^2}=
{\big[1+Y^{a+1}(l)\big]\big[1+Y^{a-1}(l)\big]
 \over
\big[1+Y^a(l)\big]^2}
\label{SYS}
\ee
If we go to the large N limit and introduce the continuous variables
(\ref{PAR}) we obtain from the previous equation a
simplified version of the differential HE (\ref{DE}):
\be
\partial_\lambda^2 \log Y = \partial_\nu^2 \log (1+Y)
\label{SDE}
\ee
which is again integrable, as it is obvious from the above determinant
formulas. In the last equations we changed a bit the definition of the
function $Y$ rescaling the variables by $1/N$.

Take first a simple example of the potential $V(x)=|x|$. The direct
calculation of (\ref{D1MM}) gives:
\be
t_a(l)= \prod_{k=1}^a(l-a+k)!(k-1)!
\label{MPT}
\ee
which yields $Y_a(l)$ as
\be
Y_a(l)={ a \over l-a+1 }
\label{MPY}
\ee
or, in the large N limit:
\be
Y_\nu(\lambda)={ \nu \over \lambda-\nu }
\label{MPYC}
\ee
which perfectly satisfies the eq. (\ref{SDE}).

To find a general (up to some comments which will follow) solution of
the eq. (\ref{SDE}) we  just have to apply the well known
formulas for the saddle point approximation in the one matrix model
with the potential which is now $V(x)+(\lambda-\nu) \log x$. Omitting
the standard calculations (see for example [\cite{GM}]) we give the result: the
function $Y_\nu(\lambda)$ obeys the following system of equations on
$Y$ and an intermediate variable $S$:
\be
\int_{-1}^1 {du \over \pi} {V'(\sqrt{Y}u+S) \over \sqrt{1-u^2} } =
{\lambda - \nu \over \sqrt{S^2 -4 Y} }
\label{SSH1}
\ee
\be
\int_{-1}^1 {du \over \pi}
{(\sqrt{Y}u+S) V'(\sqrt{Y}u+S) \over \sqrt{1-u^2} } = \lambda + \nu
\label{SSH2}
\ee
So we have obtained the solution of the eq. (\ref{SDE}) in terms of a
system of ordinary equations. For example, for a polynomial potential
the equations will become algebraic. In general they are functional.

Not every potential is compatible with this solution. We restricted
ourselves to the so called one cut solution implying the existence of
one classically stable well in the potential. This restricts our solution
to some parametrically general but still limited class of solutions
of (\ref{SDE}). The generalization to the multi-cut solution which is
straightforward should in principle yield the most general solution of
(\ref{SDE}).

The equations (\ref{SSH1}-\ref{SSH2}) look like the characteristics method of
solution of the eq. (\ref{SDE}). In the next sections we shall see to
what extent we can generalize it to the full differential HE (\ref{HE}).

\subsection{Gaussian Chain Solution of HE}

Now we shell consider another particular example of solution of the HE
(\ref{HE}) by restricting all the potentials in the matrix chain to be
Gaussian. Then all the integrals over $\phi(1),...,\phi(T)$ in
(\ref{MC}) can be easily performed and we are left in (\ref{TAU}) with
the following two matrix integral over the endpoint variables:
\be
Z_a(l,m)= \int d^{a^2} \phi \int d^{a^2} \phi' \ (\det\phi)^l
\exp - tr({1 \over 2} \phi^2+{1 \over 2}\phi'^2 -c \phi \phi')
(\det\phi')^m
\label{TAUM}
\ee

If we do the same with the continuous quantum mechanical integral
(\ref{MQM}) we arrive (up to a trivial coefficient typical for the
Green`s function of the harmonic oscillator) at the same two matrix
integral, with $c=1/\cosh(\omega T)$, where $\omega$ is the frequency
of the corresponding oscillator.

An important comment is in order. Although this partition function
satisfies the HE for three variables it should a little bit modified
for finite N: note that $Z_a(l,m)$ in (\ref{TAUM}) after passing to
eigenvalues splits into the product of two determinants corresponding
to $i,j$ both even or both odd (the matrix elements with different
parities of $i,j$ are zero, see the formula (\ref{DR}) with the
gaussian kernel). Each of these determinants satisfies the same
difference HE with the shifts of discrete variables by $\pm 2$ and not
by $\pm 1$. The continuous (large N) version of HE will be the same as
before.

This two-matrix model has the only complication with respect to the
ordinary one, containing usually only polynomial potentials: its
potentials contain logarithmic parts, like in the well know one matrix
Penner model. To solve it the ordinary method of orthogonal
polynomials does not look convenient. We propose here another, rather
powerful method worked out in a series of papers
[\cite{DOKA,KSWI,KSWII,KSWIII,K1,K2}]  and capable to solve some even more
sophisticated models than the present one.

First we perform  the integral over the relative ``angles'' of two
matrices in (\ref{TAUM}) by means of the character expansion [\cite{IZ}]:
\be
\int (d\Omega)_{U(N)} \exp[c \ tr(\phi\Omega^+ \phi'\Omega)] =
\sum_{0 \le h_a < \cdots < h_1 < \infty}
 {c^{\sum_k h_k-a(a-1)/2} \over \prod_k h_k!}\chi_h(\phi)\chi_h(\phi')
\label{IZ}
\ee
We dropped here some unessential overall coefficient.

Plugging this formula into the eq. (\ref{TAUM}) we encounter two
identical independent Gaussian integrals over $\phi$ and $\phi'$ with
the characters as pre-exponentials. These integrals can be calculated
(they slightly generalize the similar integrals appearing in
[\cite{IDF,KSWI,KSWII,KSWIII}] to the case of $l,m \ne 0$).
The result (again up to some unessential factor) is:
\be
\int d^{a^2}\phi e^{-{1\over 2} Tr \phi^2}  \chi_h(\phi) (\det \phi)^l=
{\prod_i (h^e_i+l-1)!! \prod_j (h^o_j+l)!!\over h^o_j!}
\Delta(h^e)\Delta(h^o)
\label{GI}
\ee
where we denote by $h^{e}(h^{o})$ the even(odd) highest weights whose
numbers should be equal. $\Delta(h)$ is the Van-der-Monde determinant
of $h$'s. We chose here $l$ to be even; for $l$ odd one only has to
exchange $h^{e}$ and $h^{o}$ in (\ref{GI}).

Putting all this together we obtain for (\ref{TAUM}) a representation
in terms of the multiple sum over $h$'s (we dropped a $h$-independent
coefficient):
\ba
Z_a(l,m)=
\sum_{\{h^e,h^o\}}\Delta^2(h^e) \Delta^2(h^o)
c^{\sum_k(h^e_k+h^o_k)} \times
\\
\nonumber
\times {\prod_i (h^e_i+l-1)!! \prod_j (h^o_j+l)!!\prod_i (h^e_i+m-1)!!
\prod_j (h^o_j+m)!! \over \prod_n h_n! }
\label{FSGP}
\ea
We chose here $l,m$ to be both even. Different parity for them is
forbidden by the symmetry $\phi \rightarrow -\phi$ of
(\ref{TAUM}). This will not be important in the large N limit.
As we see, the sums over $h^e$ and $h^o$ are decoupled
and can be calculated independently.

The method of calculation of these multiple sums in the large N limit
was proposed in [\cite{DOKA}] and further elaborated in
[\cite{KW,KSWI,KSWII,KSWIII}] and is based on the saddle point
approximation of this sum.  One introduces the resolvent function of
shifted highest weights:
\be
H(h)=\sum_{k=1}^a {1 \over h-h_k}
\label{HFN}
\ee

In what follows we change $h$ by $h/N$ (since the highest weights are
supposed to be of the order N in the large N limit).  So, in the large
N limit:
\be
H(h)=  \int_0^d d h'{ \tilde \rho(h') \over h-h'}=
H_+(h)\pm i \pi \tilde \rho(h)
\label{HFC}
\ee
where $H_+(h)$ is the symmetric part of the function $H(h)$ on the cut
defined by the distribution of $h$'s and $\tilde \rho(h)$ is the
density of $h$'s along this cut. In the large N limit we can calculate the
multiple sum by the saddle point method.  The saddle point condition
 defines the most probable Young tableau shaped by the density
$\tilde \rho(h)$:
\be
\barint_0^d\ dh'\ {\tilde \rho(h') \over h-h'} =
- {1 \over 2}\ln \left({c^2(h+\lambda)(h+\mu)\over h^2}\right)
\label{SPET}
\ee

One has to remember that a part of the most probable Young tableau is
in general empty (some of the highest weight components $m_k$ are
equal to zero, see [\cite{DOKA,KSWI}] for the details). So, the function
$\tilde \rho(h)$ is equal to one on the interval $(0,b)$ and to some
nontrivial function $\rho(h)$ on the interval $(b,d)$. This yields,
instead of (\ref{SPET}), the equation:
\be
\barint_b^d\ dh'\ {\rho(h') \over h-h'} =
- {1 \over 2}\ln \left({c^2(h+\lambda)(h+\mu)\over (h-b)^2}\right)
\label{SPE}
\ee
This linear integral equation has a one-cut solution:
\ba
H(h)=
\ln \left[ c^{-1}(d-b) h  \right]   \\
\nonumber
-{1\over 2}\ln
\left[(b+d+2\lambda)h-(b+d)\lambda-2db+
2\sqrt{(d+\lambda)(b+\lambda)(h-d)(h-b)}\right]   \\
\nonumber
-{1\over 2}\ln
\left[(b+d+2\mu)h-(b+d)\mu-2db+2\sqrt{(d+\mu)(b+\mu)(h-d)(h-b)}\right]
\label{SOLH}
\ea
To fix $d$ and $b$ we should recall the asymptotic of $H(h)$ with
respect to large $h$:
\be
H(h) = \nu/h + <h>/h^2 + 0(1/h^3)
\label{AS}
\ee
following from (\ref{HFN}).
Here $<h>$ is the average shifted  highest weight in
the most probable Young tableau. Expanding $H(h)$ in $1/h$ up to the terms
$O(1/h^2)$ we obtain a system of equations defining $d$ and $b$:
\ba
\left[2\lambda+d+b+2\sqrt{(d+\lambda)(b+\lambda)}\right]
\left[2\mu+d+b+2\sqrt{(d+\mu)(b+\mu)}\right]
= c^{-2}(d-b)^2   \\
\sqrt{(d+\lambda)(b+\lambda)}+\sqrt{(d+\mu)(b+\mu)}= 2\nu+\lambda+\mu
\label{AB}
\ea

{}From  (\ref{FSGP}) we deduce the following formula for the
solution of  the continuous HE (\ref{NHE}):
\be
F''_{\lambda \mu}= {1 \over 4} <<\ln (h+\lambda ),\ln (h+ \mu )>>
\label{DDF}
\ee
where by $<<A,B>>$ we denoted the connected average of any two
$h$-dependent functions $A$ and $B$. Note that this average has a
finite large N limit, as it should be.

This solution can be brought into a more explicit form: the explicite
formula for such correlators in the one matrix models was given in
[\cite{AJM,IK}] for $W(z,z')=<<{1 \over z-h )},{1 \over (z'- h )}>>$:
\be
W(z,z')=
{1 \over 2} \left[ \sqrt{{(z-d)(z'-b) \over (z'-d)(z-b)}} +
\sqrt{{(z'-d)(z-b) \over (z-d)(z'-b)}} -{1 \over (z-z')^2 } \right]
\label{CORR}
\ee
Integrating it in $z$ and $z'$ and putting $z=-\lambda$, $z'=-\mu$ we
obtain a rather explicit solution of differential HE (\ref{NHE})
\be
F''_{\lambda \mu}={1 \over 8} [g(\lambda,b,d) g(\mu,d,b)
+g(\lambda,d,b) g(\mu,b,d)+\ln(\lambda+\mu)]
\label{FEX}
\ee
where
\be
g(z,b,d)= \sqrt{(z+d)(z+b)}+ {d-b \over 2} \sqrt{{z+d\over z+b}}
\ln[2\sqrt{(z+d)(z+b)}-2 z -d-b]
\label{GFU}
\ee
and $d$,$b$ are defined by  the eqs. (\ref{AB}).

In the next section we will give the solution of (\ref{DE}) in the
case of a general time-independent potential $V(x)$. We will also
reduce the search for the most general solution of continuous Hirota
equation (\ref{DE}) defined by the time dependent potential $V(x,t)$
to a simpler problem.

\section{ COLLECTIVE FIELD METHOD \\
FOR THE DIFFERENTIAL HE}

Now let us briefly describe how our method works in the case of a the
general potential in the matrix quantum mechanics defined by
(\ref{MQM}). In this case we can apply the collective coordinate
method of Jevicki and Sakita [\cite{JS}] which is valid in the large N
limit and can be applied to the non-stationary saddle point solutions
which are needed in our case. The details of this approach can be
found in [\cite{JS,DJ,AJ,AM}]. We will use the results of it and apply
them to our case.

In terms of this method the effective action for (\ref{MQM}) can be
written for the density $\rho(x,t)$ of  eigenvalues $x_k$ and its
conjugate momentum $P(x,t)$ developing in time as:
\be
S_{eff}[\rho,P]= \int dx \int_0^T dt \big[ \dot \rho P
-{1 \over 2} \rho P'^2 +{\pi^2 \over 6}\rho^3
+\rho  V(x,t)\big]
\label{SEFF}
\ee
reflecting our specific boundary conditions at the ends of the
interval $(0,T)$ following from (\ref{TAU}).

The equations of motion corresponding to the action (\ref{SEFF}) are:
\be
 \dot \rho+ \partial_x(P' \rho)=0
\label{EOM1}
\ee
\be
\dot P+ {1 \over 2} P'^2 =
 {\pi^2 \over 2}\rho^2 +  V(x,t)
\label{EOM2}
\ee
Differentiating the second one in $x$ and defining the
function $f(x,t)=  P' +i\pi \rho$ we rewrite this system as
only one forced Hopf equation on this complex function:
\be
 \partial_t f+ f \partial_x f = -V'(x,t)
\label{SEOM}
\ee
 We have to impose at any moment, say, at $t=0$
the normalization condition
\be
\int dx \rho(x,0)=\nu
\label{NC}
\ee
Then it will be true at any $t$ due to the condition (\ref{EOM1}).

We have excluded the end points of the interval $(0,T)$ in the last
equation. The logarithmic potentials  and the two
Van-der-Monde determinants left at the ends of the interval can be
taken into account as the boundary conditions:
\be
{\it Re}f(x,0)={\lambda \over x}+\barint_{b(0)}^{d(0)}\ dx'\ { \rho(x',0)
\over x-x'}, \ \ b(0) \le x \le d(0)
\label{BCF1}
\ee
\be
{\it Re}f(x,T)=-{\mu \over x}-\barint_{b(T)}^{d(T)}\ dx'\ { \rho(x',T)
\over x-x'}, \ \ b(T) \le x \le d(T)
\label{BCF2}
\ee
or, introducing the resolvent: $R_{\pm}(x,t)= \barint_{b(t)}^{d(t)}\
dx'\ { \rho(x',t) \over x-x'} \pm i\pi \rho(x,t)$,
\be
f(x,0)={\lambda \over x}+R_+(x,0)
\label{BK1}
\ee
\be
f(x,T)=-{\mu \over x}-R_-(x,T)
\label{BK2}
\ee

Hence we reduced the problem of the virtually general (since it is
parameterized by a general potential depending on 2 variables) solution
of the continuous HE (\ref{HE}) on the function of 3 variables $\nu$,
$\lambda$ et $\mu$ to the solution of the differential equation of the
first order (\ref{SEOM}) (assuming that the complex function $f$ is
analytical in their variables) supplemented by the boundary conditions
(\ref{BCF1}-\ref{BCF2}). The variables $\nu$, $\lambda$ et $\mu$
appear here only as fixed parameters.

It is still quite complicated, although simpler than the original
problem and, may be, physically more transparent. We don't know how we
could simplify it further in general case.  So let us consider an
interesting particular example of the time independent potential
$V(x)$. In that case the forced Hopf equation
\be
 \partial_t f+ f\partial_x f = -V'(x)
\label{XEOM}
\ee
becomes completely integrable by the characteristics method
\footnote{I am grateful to A. Matytsin for the explanation of this
method and its application to the forced Hopf equation}.

We will choose $\lambda= \bar \mu$, which does not look as restriction
if we assume the analyticity in $\lambda$ and $\mu$ Then due to the
time reversal symmetry we have $f(x,T)=-\bar f(x,0)$.  So we can say
that also $\rho(x,T)= \rho(x,0)$. Hence only one of these two boundary
conditions is independent.

The first part of the problem is to find the solution of
eq. (\ref{XEOM}) with  fixed endpoint density $\rho(x,T)=
\rho(x,0)$. The result can be formulated as the following equation on
the functions already at the endpoints
\be
T = \int_x^{G(x)} {dy \over \sqrt{2(g_0(x)-V(y))}}
\label{INT}
\ee
where $G(x)=x_T(g_T(x))$ and $x_T(g)$ is the function to be found by
solution of
\be
g(x_T)={1\over 2} f^2(x_T,T) + V(x_T)
\label{DEF1}
\ee
Note that the function  $\bar G(x)=x_0(g_0(x))$ defined by the solution of
\be
g={1\over 2} f^2(x_0,0) + V(x_0)
\label{DEF2}
\ee
is the functional inverse of the function $G(x)$ itself which gives
the equation of A. Matytsin [\cite{AM}]:
\be
G(\bar G(x))=x
\label{INV}
\ee
Although the dynamics of the forced Hopf equation is summarized by the
relation (\ref{INT}) the last equation leads to a strong constraint on
the analytical structure of the function $G(x)$.

Once we found the functional $f_{\rho(x,0)}(x,0)$ as the solution of
eqs. (\ref{INT}-\ref{INV}) we have to much it with our boundary
conditions:
\be
{\it Re} f_{\rho(x,0)}(x,0)=
{\lambda \over x}+\barint_{b(0)}^{d(0)}\ dx'\ { \rho(x',0)
\over x-x'}, \ \ b(0) \le x \le d(0)
\label{BCF}
\ee

With a given $V(x)$ this defines the end-point density
$\rho(x,0)$ which is the only non-trivial information we  need to
find the quantity (\ref{TAU}). Probably it is convenient to represent this
boundary condition as a condition on the large $x$ asymptotic of
$f(x,0)$:
\be
f(x,0) \rightarrow_{x \rightarrow \infty } {\nu+ \lambda \over x}
\label{ASM}
\ee

We thus reduced the solution of the continuous HE (\ref{NHE}) to some
simpler functional problem in a particular but rather representative
case of the time independent potential. The solution with the harmonic
oscillator potential obtained in the previous section should be also
reproducible by this method.

We can also use these equation to produce  more explicit solutions
of the HE (\ref{NHE}) by the method proposed in [\cite{AM}]: one
chooses two conjugated roots $G(x)$ and $\bar G(x)$ of an algebraic
equation $x(G)=0$, where $x(G)$ is some polynomial. These two roots
satisfy by construction the eq. (\ref{INV}). Then one has to plug
them into (\ref{INT}) and solve it as an integral equation for $V(x)$.
Of course the choice should be limited by the boundary conditions
(\ref{BCF}) or (\ref{ASM}).

It would be interesting to analyze  the case of the inverted oscillator
potential corresponding to the 1+1 dimensional non-critical string
theory. But this question lies beyond the scope of this paper.

\section{CONCLUSIONS AND PROSPECTS}

In this paper we proposed a matrix model representation for the
solution of the general difference Hirota equation. For its continuous
analog of the differential HE of second order on three variables the
solution is represented by the large N limit of the corresponding
matrix chain. It gives an effective framework for solving the
continuous HE in a rather explicit way, at least for some particular
cases. In particular, the problem can be reduced to the forced Hopf
equation with specific boundary conditions.

Many things remain to be understood. First of all it is not clear how
to find the solutions satisfying the boundary conditions of various
1+1 dimensional quantum field theories solvable by Bethe
ansatz. Especially how to choose the matrix potentials to get the
relativistic spectrum for the physical particles and to fix the
symmetries of original models. Another question: is there some
physical interpretation in terms of these integrable theories of the
time variable $t$ and of the eigenvalue variable $x$ in our matrix
representation similar to the non-critical (1+1) dimensional string
theory (where these variables describe the target space of the string
[\cite{DJ,POL}])?  It might be for example that the time can be
considered as the physical space dimension of the corresponding
integrable theory.  But it remains to be proved. We hope that the
formalism proposed here at least sets a convenient framework for the
attempts to find the physically interesting solutions.

A more formal use of our method might be the search for new solitonic
solutions for the well known integrable equations. For example, the
Toda equation $\partial_\nu^2 F =
\partial_\lambda \partial_\mu \exp F$ is just a particular
limit of the continuous HE (\ref{DE}). To our knowledge, the solitonic
solutions to this equation are not yet found.

Another interesting question is how the double scaling limit in matrix
models is related to HE? For example, how to find the corresponding
solution for the inverted harmonic oscillator giving the description
of the (1+1) dimensional string field theory. To answer this question
as well as many others we have to learn how to deal with the
non-stationary forced Hopf equation with our specific boundary
conditions. The methods worked out in the papers
[\cite{AM,MAT1,MAT2,WAD}] could be useful for that.

\begin{center}
{\bf Acknowledgments}
\end{center}

I would like to thank P. Wiegmann, I. Kostov, A. Matytsin and
especially P. Zinn-Justin for many valuable discussions. I am also
grateful to the Theory Division  of CERN where
a part of this work was done for the kind hospitality.

\end{document}